\documentclass[prb,twocolumn,amsmath,amssymb,showpacs]{revtex4}

\usepackage{graphicx}
\usepackage{dcolumn}
\usepackage{bm}
\usepackage{mathrsfs}
\usepackage{appendix}
\usepackage{bbm}
\usepackage{color}

\def \vS {{\bf S}}
\def \vR {{\bf R}}
\def \ve {{\bf e}}
\def \vQ {{\bf Q}}
\def \vq {{\bf q}}
\def \vh {{\bf h}}
\def \mb {\mu_{\rm B}}
\def \sgn {{\rm sgn}}
\def \pin {{\bf P}^{\rm ind} }
\def \xp {{\bf x}^{\prime }}
\def \yp {{\bf y}^{\prime }}
\def \zp {{\bf z}^{\prime }}
\def \xx {{\bf x}}
\def \yy {{\bf y}}
\def \zz {{\bf z}}

\def \vM {{\bf M}}
\def \vP {{\bf P}}
\def \BF {{\rm BiFeO$_3$} }
\def \BP {{\rm BiFeO$_3$}}
\def \TC {T_c}
\def \TN {T_{\rm N}}
\def \uX {{\underline X}}
\def \vk {{\bf k}}
\def \vv {{\bf v}}
\def \ww {{\bf w}}
\def \xq {x^{\prime}}
\def \yq {y^{\prime}}
\def \zq {z^{\prime}}
\def \ds {\displaystyle }
\def \alp {\alpha^{\prime}}
\def \cdp {\chi^{\prime \prime }}
\def \Ui {U^{-1\, r}}

\begin{document}

\title{The Spin State and Spectroscopic Modes of Multiferroic BiFeO$_3$}

\author{Randy S. Fishman,$^1$ Jason T. Haraldsen,$^{2,3}$ Nobuo Furukawa,$^4$ and Shin Miyahara$^5$}

\affiliation{$^1$Materials Science and Technology Division, Oak Ridge National Laboratory, Oak Ridge, Tennessee 37831, USA}
\affiliation{$^2$Theoretical Division, Los Alamos National Laboratory, Los Alamos, New Mexico 87545, USA}
\affiliation{$^3$Center for Integrated Nanotechnologies, Los Alamos National Laboratory, Los Alamos, New Mexico 87545, USA}
\affiliation{$^4$Department of Physics and Mathematics, Aoyama Gakuin University, Sagamihara, Kanagawa 229-8558, Japan}
\affiliation{$^5$Asia Pacific Center for Theoretical Physics, Pohang University of Science and Technology, Pohang, Gyeongbuk, 790-784, Korea}

\date{\today}

\begin{abstract}

Spectroscopic modes provide the most sensitive probe of the very weak interactions responsible for the properties of the 
long-wavelength cycloid in the multiferroic phase of \BF below $\TN \approx 640$ K.  Three of the four modes measured by THz and Raman spectroscopies were 
recently identified using a simple microscopic model.  While a Dzyaloshinskii-Moriya (DM) interaction $D$ along $[-1,2,-1]$ induces the cycloid with wavevector 
$(2\pi /a)(0.5+\delta ,0.5, 0.5-\delta )$ ($\delta \approx 0.0045$), easy-axis anisotropy $K$ along the $[1,1,1]$ direction of the electric polarization ${\bf P}$ 
induces higher harmonics of the cycloid, which split the $\Psi_1$ modes at 2.49 and 2.67 meV and activate the $\Phi_2$ mode at 3.38 meV.
However, that model could not explain the observed low-frequency mode at about 2.17 meV.  We now demonstrate that an additional 
DM interaction $D'$ along $[1,1,1]$ not only produces the observed weak ferromagnetic moment of the high-field phase above 18 T but also 
activates the spectroscopic matrix elements of the nearly-degenerate, low-frequency $\Psi_0$ and $\Phi_1$ modes, although their 
scattering intensities remain extremely weak.  Even in the absence of easy-axis anisotropy, $D'$ produces cycloidal harmonics 
that split $\Psi_1 $ and activate $\Phi_2$.  However, the observed mode frequencies and selection rules require that both $D'$ and $K$ 
are nonzero.  This work also resolves an earlier disagreement between spectroscopic and inelastic neutron-scattering measurements.

\end{abstract}

\pacs{75.25.-j, 75.30.Ds, 78.30.-j, 75.50.Ee}

\maketitle

\section{Introduction}

As the only known room-temperature multiferroic, \BF continues to attract a great deal of attention.
Multiferroic materials offer the tantalizing prospect of controlling magnetic properties with electric fields or 
electric polarizations with magnetic fields \cite{eerenstein06}.  Although the ferroelectric transition temperature \cite{teague70} $\TC \approx 1100$ K 
of \BF is far higher than its N\'eel temperature \cite{sosnowska82, lebeugle08, slee08a} $\TN \approx 640$ K,
the electric polarization $\vP $ is enhanced by its coupling to the long-wavelength cycloid below $\TN $ [\onlinecite{park11}].
As a result, the magnetic domain distribution below $\TN $ can be manipulated by an electric field \cite{lebeugle08,slee08a,slee08b}. 

Before \BF can be used in technological applications, however, it is essential to understand the microscopic mechanisms and 
interactions responsible for its magnetic behavior.  At frequencies above a few meV up to about 70 meV, the spin-wave (SW) spectrum 
of \BF has been used \cite{jeong12, matsuda12} to determine the nearest-neighbor and next-nearest neighbor exchange interactions 
$J_1\approx -4.5$ meV and $J_2 \approx -0.2$ meV between the $S=5/2$ Fe$^{3+}$ spins \cite{temp} on a pseudo-cubic lattice 
with lattice constant $a\approx 3.96\, \AA $.  As shown in Fig.1(a), $J_1$ is the antiferromagnetic (AF) interaction between 
spins on neighboring $(1,1,1)$ planes separated by $c= a/\sqrt{3}$ while $J_2$ is the AF interaction between neighboring 
spins on each hexagonal layer.  

Below $\TN $, a long-wavelength cycloid with wavevector $\vQ = (2\pi /a)[0.5+\delta ,0.5, 0.5-\delta ]$ ($\delta \approx 0.0045$) 
\cite{sosnowska82, rama11a, herrero10, sosnowska11} is produced by the Dzyaloshinskii-Moriya (DM) interaction 
${\bf D}=D\yp $ along $\yp = [-1,2,-1]$ (all unit vectors are assumed normalized to one).  As shown in Fig.1(b), 
the spins of the cycloid lie predominantly in the $(-1,2,-1)$ plane normal to $\yp $.

\begin{figure}
\includegraphics[width=8.5cm]{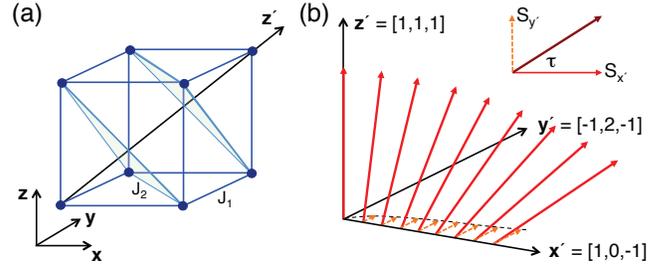}
\caption{(Color online) (a) The pseudo-cubic cell with exchange interactions $J_1$ and $J_2$ as well as the 
polarization direction $\zp $ cutting through two hexagonal planes.  (b) For domain 1, a schematic of the spins along the $\xq $ axis 
showing their rotation about $\yp $.  Due to the DM interaction ${\bf D}'=D' \zp$, spins rotate by 
$\tau $ about $\zp $ in the $\xq \yq$ plane.
}
\end{figure}

Whereas the high-frequency portion of the SW spectrum determines the Heisenberg exchange interactions, the low-frequency modes 
measured by THz \cite{talbayev11, nagelun} and Raman \cite{cazayous08, singh08, rovillain09} spectroscopies can be used to determine the 
small microscopic interactions that control the cycloid.  Four modes have been detected at frequencies \cite{temp} of
2.17, 2.49, 2.67, and 3.35 meV.  By comparison, a model with the single DM interaction ${\bf D}$ only produces \cite{fishman12} a 
single spectroscopically-active mode labeled $\Psi_1$ at about 2.37 meV.

A more realistic model \cite{fishman12, sosnowska95} also contains the easy-axis anisotropy $K$ along $\zp = [1,1,1]$, parallel to 
the electric polarization $\vP $.  When $K > 0 $, $\Psi_1$ splits into two and $\Phi_2$ at 3.38 meV is 
activated \cite{fishman12}.  Although this model successfully described the upper three spectroscopic modes, with predicted 
frequencies very close to the measured frequencies, it failed to explain the low-frequency 2.17 mode.  In addition, it provides 
conflicting estimates for $K$ based on spectroscopic and inelastic neutron-scattering measurements.

Several authors \cite{kadomtseva04, ed05, pyatakov09, ohoyama11} have examined the effects of another DM interaction 
${\bf D'}= D'\zp $ between neighboring hexagonal layers.  For a G-type AF, $D'$ produces a weak ferromagnetic moment 
along $\yp $ due to the canting of the uniform moments on each hexagonal plane.  The moment ${\bf M}_0 = 2\mb S_0\yp \approx 0.03 \mb \yp $ 
was subsequently observed in the metamagnetic phase \cite{tokunaga10, park11} above 18 T.  Below 18 T, $D'$ was predicted
\cite{pyatakov09} to induce an oscillatory component of the cycloid along $\yp $, which has 
recently been confirmed by neutron-scattering measurements \cite{rama11b}.

Based on a model that includes both $D$ and $D'$ in addition to the easy-axis anisotropy $K$, we evaluate the spin state and 
spectroscopic modes of \BP .  Even when $K=0$, $D'$ induces higher harmonics of the cycloid that split $\Psi_1$ 
and activate $\Phi_2$.  More remarkably, $D'$ activates $\Psi_0$ and $\Phi_1$ at the cycloidal wavevector.   

We believe that these nearly-degenerate modes are responsible for the low-frequency 2.17 meV peak observed in 
spectroscopy measurements.  Although a model with $K=0$ can produce four spectroscopic modes, the $\Psi_1$ selection 
rules are reversed and their mode frequencies are too small.  Therefore, both $D'$ and $K$ are required to explain the experimental 
measurements.  With $D' \approx 0.054$ meV, corresponding to the observed value \cite{tokunaga10, park11} $S_0=0.015$, 
we estimate that $D\approx 0.11$ meV and $K \approx 0.0035$ meV, which also provide a good description of
inelastic neutron-scattering measurements \cite{matsuda12} below 5 meV.

This paper is divided into seven sections.  Section II constructs the spin state of \BP.  Section III evaluates the spin dynamics of that state,
Section IV evaluates the spectroscopic modes of that state, and Section V discusses the selection rules for those modes.
Section VI discusses the inelastic neutron-scattering spectrum for the low-frequency modes.  Section VII contains a brief conclusion.  
Results for the SW intensities are provided in Appendix A.  The polarization and magnetic matrix elements are provided in Appendix B.

\section{Spin State}

With $\vP =P\zp $, the three magnetic domains have cycloidal wavevectors $\vQ = (2\pi /a )[0.5+\delta , 0.5, 0.5-\delta ]$ (domain 1),
$(2\pi /a )[0.5, 0.5+\delta , 0.5-\delta ]$ (domain 2), or $(2\pi /a )[0.5+\delta , 0.5-\delta ,0.5]$ (domain 3).   By contrast, the G-type AF 
stabilized by a magnetic field \cite{tokunaga10, park11}, doping \cite{chen12}, or in thin films \cite{bai05} has wavevector $(2\pi /a)[0.5,0.5,0.5]$.
In our discussion of the selection rules governing the spectroscopic modes in Section V, we will assume that all three domains are equally populated.   
Since the spin state and dynamics are the same for all three domains, we now concentrate on domain 1 with $\xp = [1,0,-1]$ and $\yp =[-1,2,-1]$, 
as shown in Fig.1(b).

The spin state and SW excitations of \BF are evaluated from the Hamiltonian
\begin{eqnarray}
&&H = -J_1\sum_{\langle i,j\rangle }\mbox{\boldmath $S$}_i\cdot\mbox{\boldmath $S$}_j -J_2\sum_{\langle i,j \rangle'} \mbox{\boldmath $S$}_i\cdot\mbox{\boldmath $S$}_j
-K\sum_i {S_{i\zq }}^2
\nonumber \\
&&-{D\, \sum}_{\vR_j=\vR_i + a(\xx -\zz )} \,\yp \cdot (\mbox{\boldmath $S$}_i\times\mbox{\boldmath $S$}_j) \\
&& - {D'\, \sum}_{\vR_j=\vR_i + a\xx, a\yy, a\zz } \, (-1)^{R_{i\zq } /c} \,\zp \cdot  (\mbox{\boldmath $S$}_{i}\times\mbox{\boldmath $S$}_j).
\nonumber
\label{Ham}
\end{eqnarray}
The first and second exchange terms contain sums $\langle i,j\rangle $ and $\langle i,j\rangle'$ over nearest and next-nearest neighbors on the 
pseudo-cubic lattice.  The third term arises from the easy-axis anisotropy along $\zp $ and the fourth term from the DM interaction with ${\bf D}=D\yp $.

Compared to the model for \BF introduced in Ref.[\onlinecite{sosnowska95}] and studied in our earlier work \cite{fishman12}, 
$H$ adds the DM interaction ${\bf D'}=D' \zp $.  This term alternates in sign with increasing $\zq $:  
$(-1)^{R_{i\zq } /c}$ changes sign from layer $n$ to layer $n+1$ so the DM interaction $(-1)^{R_{i\zq } /c}D'$
between layers $n$ and $n+1$ has opposite sign to the DM interaction between layers $n+1$ and $n+2$.
Hence, the DM interaction ${\bf D'}$ has the same wavevector $(2\pi /a)[0.5,0.5,0.5]$ as a G-type AF.

Because $\delta \approx 1/222$, a unit cell containing $M = 222$ sites within each of two neighboring $(1,1,1)$ planes is used to characterize the distorted cycloid.
In zero magnetic field, the cycloid can be expanded in odd harmonics  \cite{zhi96, fishman10} of the fundamental wavevector $\vQ $ 
(even harmonics are also required in non-zero fields).  If $S_{\yq }(\vR)$ is proportional to $S_{\xq }(\vR)$, then
\begin{eqnarray}
\label{dcx}
S_{\xq }(\vR )&& = (-1)^{R_{\zq }/c} \sqrt{1-\kappa^2} \sqrt{S^2-S_{\zq }(\vR )^2}  \nonumber \\
&& \sgn \bigl(\sin (2\pi \delta R_{\xq } /a  ) \bigr),\\
\label{dcy}
S_{\yq }(\vR ) &&= \kappa \sqrt{S^2-S_{\zq }(\vR )^2} \, \sgn \bigl(\sin (2\pi \delta R_{\xq } /a ) \bigr),
\\
\label{dcz}
S_{\zq }(\vR )&&= (-1)^{R_{\zq } /c } S\nonumber \\
&& \sum_{m=0}^{\infty } C_{2m+1} \cos \bigl( (2m+1)2\pi \delta R_{\xq } /a \bigr).
\end{eqnarray}
Odd-order coefficients $C_{2m+1}$ in $S_{\zq }(\vR)$ satisfy $\sum_{m=0}^{\infty }C_{2m+1} =1$.
Although $S_{\yq }(\vR)$ (unlike $S_{\xq }(\vR)$ and $S_{\zq }(\vR)$) does not change sign from one layer to the next,
the average value of $S_{\yq }(\vR) $ vanishes and there is no net moment in any direction.
The ratio $S_{\yq }(\vR)/S_{\xq }(\vR)$ has magnitude $\kappa  /\sqrt{1-\kappa^2}$, which is proportional to $\vert D'/ J_1\vert \ll 1$.
Hence, the tilting angle $\tau $ indicated in Fig.1(b) satisfies the relation $\tan \tau = \kappa /\sqrt{1-\kappa^2} \approx \kappa $.
Although the cycloid remains coplanar for each hexagonal layer, the cycloidal planes rotate by $2\tau $ from one layer to the next.

The parameters of the spin state are evaluated by minimizing the energy $E=\langle H \rangle $ in a unit cell $\xq \yq \zq$ of dimensions 15,000$a \times a \times 2c$
containing two $(1,1,1)$ layers.  Open boundary conditions are employed along the $\xq $ direction.   
With the exchange interactions $J_1 = -4.5$ meV and $J_2=-0.2$ meV fixed
at the values required to describe the SW spectrum \cite{jeong12, matsuda12} at high frequencies, 
the four variational parameters are $\delta $, $\kappa $, $C_3$, and $C_5$.   
A solution with $\delta = 1/222$ is obtained by varying the DM interaction $D$ for fixed $K$.  After minimizing the energy, 
we verify that the corresponding spin state provides at least a metastable minimum by checking that the classical forces on each spin vanish.

With a magnetic field oriented along $\zp $, the metamagnetic state observed \cite{tokunaga10, park11} above 18 T can be written
\begin{eqnarray}
\vS_1 &=&  S\bigl(\cos \theta \cos \phi , \cos \theta \sin \phi , \sin \theta \bigr), \\
\vS_2 &=&  S\bigl(-\cos \theta \cos \phi , \cos \theta \sin \phi , \sin \theta \bigr),
\end{eqnarray}
for $R_{\zq } = 2m c$ and $(2m+1)c$, respectively.
Extrapolating to zero field with $\theta = 0$, we obtain
$\tan 2\phi = D'/J_1$.   Hence, the weak ferromagnetic moment of the metamagnetic phase is 
\begin{equation}
M_0 = 2\mu_B S_0= 2\mb S\sin \phi \approx \frac{\mb S D'}{J_1},
\end{equation}
independent of $D$, $K$, and $J_2$.
Using $J_1= -4.5$ meV and the experimental result \cite{tokunaga10, park11} $S_0=0.015$, we estimate that $\vert D'\vert = 0.054$ meV,
which is is slightly larger than the estimate $\vert D'\vert = 0.046$ meV provided in Ref.[\onlinecite{ohoyama11}].

For the distorted cycloid given by Eqs.(\ref{dcx}-\ref{dcz}), it is straightforward to show that if $\delta \ll 1$, then $\kappa \approx D'/2J_1$.
Therefore, the {\it maximum} cycloidal spin $\vert S_{\yq }(\vR )\vert $ equals the weak ferromagnetic spin $S_0$ of the metamagnetic phase.  
For the tilting angle, we estimate $\tau \approx 0.34^\circ $, a bit smaller than the recent neutron-scattering \cite{rama11b} estimate of $\sim 1^\circ $.

\begin{figure}
\includegraphics[width=8cm]{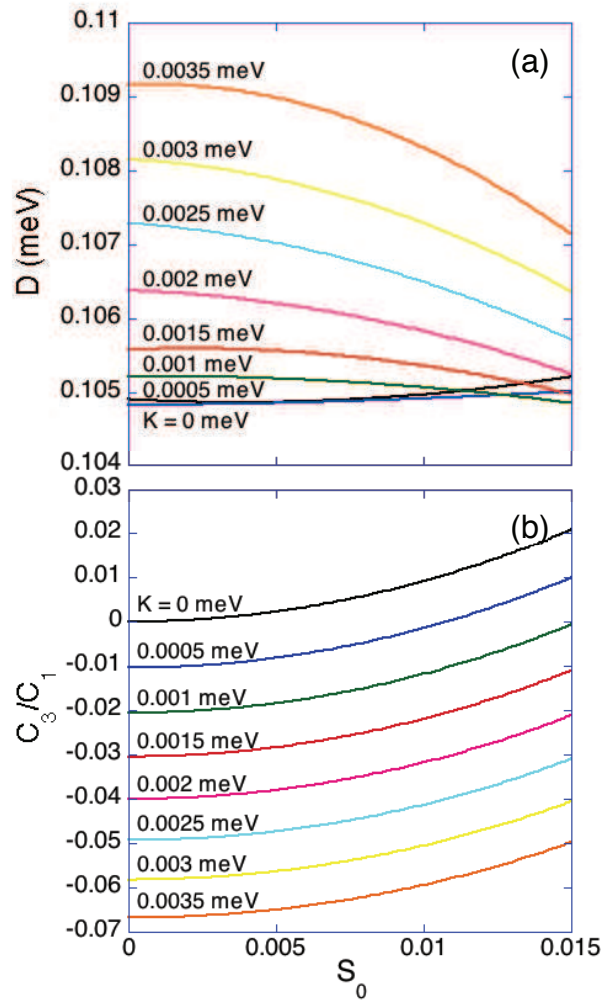}
\caption{(Color online) (a) The DM interaction $D$ and (b) the ratio of harmonics $C_3/C_1$ versus $S_0$ for several values of $K$.
}
\end{figure}

In Fig.2(a), we plot the DM interaction $D$ versus $S_0$ for several values of the anisotropy $K$ ranging from 0 to 0.0035 meV.
For $K=0$ and 0.0005 meV, $D$ increases slightly with $S_0$.  But for $K \ge 0.001$ meV,
$D$ decreases with $S_0$.  Nevertheless, the variation of $D$ with $S_0$ is rather modest.

By contrast, the higher harmonics of the cycloid exhibit a much stronger variation with $S_0$.  Fig.2(b) reveals that
the ratio $C_3/C_1$ increases with $S_0$ for all $K$.  Since $C_1 = 1-\sum_{n=1}C_{2n+1} $ and $\vert C_5\vert \ll \vert C_3 \vert $,
$C_1 \approx 1 - C_3$ and $C_3/C_1 \approx C_3(1+ C_3)$.  For $K=0$ and $S_0 > 0$, $C_3>0$ and  
\begin{equation}
\langle {S_{i\zq }}^2 \rangle = \frac{1}{2}\sum_{n=0} \bigl( C_{2n+1}\bigr)^2 \approx \frac{1}{2}\bigl(1-2C_3\bigr) < \frac{1}{2}.
\end{equation}
Because the ${\bf D'}$ interaction energy is optimized when the spins lie in the $x'y'$ plane, 
higher harmonics favor the $z'$ nodal regions of the cycloid.  When $S_0$ is sufficiently small and $K > 0$, $C_3<0$ and 
$\langle S_{i\zq }^2 \rangle > 1/2$ so that higher harmonics favor the $\zq $ antinodal regions of the cycloid.
Experimentally, the ratio of the neutron-scattering intensity from the third to the first harmonics is given by $(C_3/C_1)^2$.

Notice that the third (and higher) harmonics can vanish for nonzero $S_0$ and $K$.  When $S_0 = 0.015$, $C_3 < 0$ when $K$ is less than about  0.001 meV and
$C_3 > 0$ when $K$ is greater than about  0.001 meV.  For $K\approx 0.001$ meV, the higher harmonics of the cycloid vanish and $\langle {S_{i\zq }}^2 \rangle =1/2$.

\section{SW Excitations}

The SW frequencies are calculated using the equations-of-motion technique for non-collinear spins outlined in Ref.[\onlinecite{haraldsen09}].
A unit cell containing $M = 222$ sites on each of two hexagonal layers is constructed to evaluate the $2M$ SW frequencies
$\omega_n(\vq )$.  SW intensities are obtained from the spin-spin correlation function defined by Eq.(\ref{ssc}) in Appendix A.  
In the absence of damping, the inelastic scattering cross section $S(\vq ,\omega )$ can be expanded as the sum over delta functions at each frequency:
\begin{equation}
S(\vq ,\omega )= \sum_{n, \alpha } \Bigl(1- (q_{\alpha }/q)^2 \Bigr) \delta \bigl( \omega - \omega_n(\vq ) \bigr) S^{(n)}_{\alpha \alpha }(\vq ). \,\,\,\,\,\,\,\,\,
\end{equation}
The amplitudes $S^{(n)}_{\alpha \alpha }(\vq )$ are evaluated using Eq.(\ref{sec}).

\begin{figure}
\includegraphics[width=8.8cm]{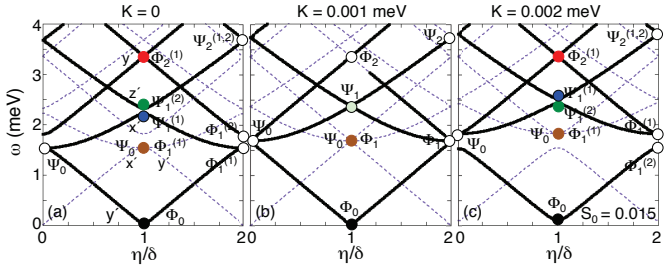}
\caption{(Color online)  The SW modes of BiFeO$_3$ versus $\eta /\delta $ for wavevector $(2\pi /a)(0.5+\eta ,0.5,0.5-\eta )$.  Dashed lines show all possible excitations and 
the solid lines show only those modes with significant intensity above a threshold value.  All three plots take $S_0=0.015$ and $D'=0.054$ meV.
$\Phi_0$ (black dot) has a very large $\yq $ MR matrix element.
The low-frequency mode (brown dot) has both $\Psi_0$ and $\Phi_1^{(1)}$ contributions with nonzero $\xq $ and $\yq $ MR matrix elements, respectively.
Whereas $\Phi_2^{(1)}$ (red) has nonzero $\yq $ MR matrix element, $\Psi_1^{(1)}$ (blue) and $\Psi_1^{(2)}$ (green) have nonzero $\xq $ and $\zq $ matrix elements,
respectively.  The EM mode with component $\yq $ coincides with $\Psi_1^{(1)}$.
}
\label{SWs}
\end{figure}

For fixed $S_0=0.015$, the SW frequencies are plotted in Fig.3 for $K=0$, 0.001, and 0.002 meV.
Although there are $2M$ modes for every wavevector $2\pi /a (0.5 + \eta , 0.5, 0.5-\eta )$, plotted by the dashed lines, only a few of those modes have any significant
intensity.  Modes with intensity above an arbitrary cutoff are plotted in the dark lines.

When $K\approx 0.001$ meV in Fig.3(b), the higher harmonics of the cycloid vanish and the SW frequencies are similar to those for 
$S_0=0$ and $K=0$ discussed in Ref.[\onlinecite{fishman12}].  In the absence of harmonics, de Sousa and Moore\cite{sousa08} labeled
the SW frequencies $\omega_n (m Q)$ ($n=1$ or 2) of a one-dimensional cycloid at multiples $m$ of the 
cycloidal wavevector $Q =2\pi \delta /a$ as $\Phi_m$ and $\Psi_m$.  Using an extended zone scheme and assuming that $\vert m\vert \delta \ll 1$, 
$\omega_n (m Q)$ can be approximated by $\Phi_m = \Phi_1 \vert m\vert $ and 
$\Psi_m = \Phi_1 \sqrt{1+m^2}$.
These relations imply that $\Phi_1=\Psi_0$, as seen in Fig.3(b), and that the $\Phi_{\pm m}$ and $\Psi_{\pm m}$ 
modes cross without repulsion at the zone center $q=Q$ and zone boundary $q=0$.  

Whether produced by the tilt $\tau $ or by the anisotropy $K$, higher odd harmonics of the cycloid 
introduce higher even harmonics in the Hamiltonian $H$.  A $2m\vQ $ potential will split the 
$\Phi_{\pm m}$ and $\Psi_{\pm m}$ modes.  As shown in Figs.3(a) and (c), the new $m=1$
eigenmodes are labeled $\Phi_1^{(1,2)}$ and $\Psi_1^{(1,2)}$.  Notice that $\Psi_0$ and $\Phi_1^{(1)}$ are nearly degenerate for all $K$.
Although too small to see in Fig.3, even $\Phi_{\pm 2}$ are split by anharmonicity.  

\section{Spectroscopic Modes}

Because the wavelength of far infrared light greatly exceeds atomic length scales, the SW modes measured by THz and Raman spectroscopies 
lie at the zone center $\vq = \vQ $ or $\eta = \delta $.  A magnetic resonance (MR) mode has nonzero matrix element
$\langle \delta \vert M_{\alpha }\vert 0\rangle $, where $\vert 0 \rangle $ is the ground state and $\vert \delta \rangle $ is an excited state 
with a single magnon of wavevector $\vQ $.   An electromagnon (EM) mode has nonzero matrix element $\langle \delta \vert P^{\rm ind}_{\alpha }\vert 0\rangle $ 
so that the induced polarization directly couples the ground state to the excited state.  

In order to evaluate the MR and EM matrix elements, we must first express the magnetic moment $\vM $ and induced polarization $\pin $
operators in terms of the spin operators $\vS_i$.   The magnetic moment $\vM= 2\mb \sum_{\vR_i} \vS_i$
contains a sum over the $2M$ unique sublattices.  In \BP , the coupling between the cycloid and electric polarization is produced 
by the inverse DM mechanism \cite{katsura05, mostovoy06, sergienko06} with induced polarization 
\begin{equation}
\pin = {\lambda \sum}_{\vR_i, \vR_j=\vR_i + \ve_{ij} } \Bigl\{ \ve_{ij}\times \bigl(\vS_i\times \vS_j\bigr) \Bigr\},
\end{equation}
where the sum is restricted to the $2M$ sublattices using periodic boundary conditions.
Within each $(1,1,1)$ plane, $\ve_{ij}= \sqrt{2}a\xp $ connects spins at sites $\vR_i$ and $\vR_j$.  
So if $\langle 0 \vert \vS_i \times \vS_j \vert 0 \rangle $ points  along $\yp $, then $\langle 0\vert  \pin \vert 0 \rangle $ points along $\zp $.

\begin{figure}
\includegraphics[width=8.5cm]{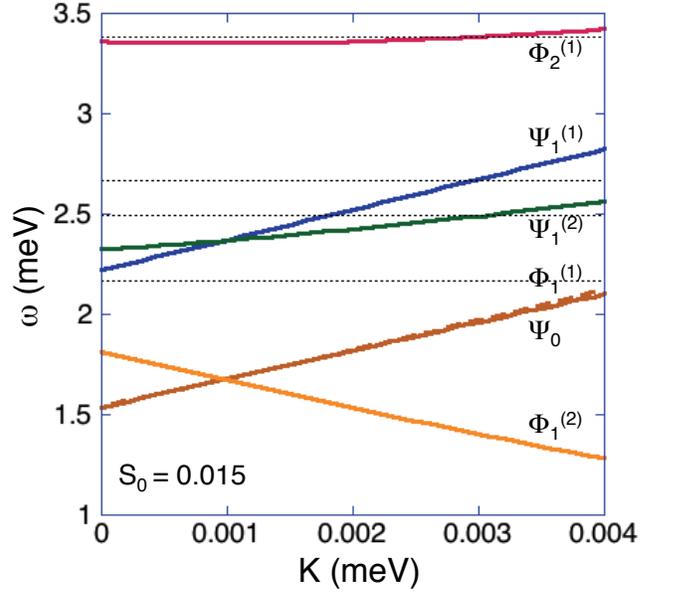}
\caption{(Color online) The evolution of the predicted modes with anisotropy $K$ taking $S_0=0.015$ and $D'=0.054$ meV.  
The horizontal dashed lines are the spectroscopic mode frequencies \cite{temp}.}
\end{figure}

Expressions for the matrix elements $\langle \delta \vert M_{\alpha }\vert 0\rangle $ and $\langle \delta \vert P^{\rm ind}_{\alpha }\vert 0\rangle $ are provided
in Appendix B.  Although there is no simple relation between the MR matrix elements and the SW intensities,
the MR and EM modes only appear at mode frequencies $n$ with $S^{(n)}_{\alp \alp }(\delta ) > 0$.  
Generally, $\Phi_n$ modes with $\langle \delta \vert M_{\yq } \vert 0 \rangle \ne 0$ also have nonzero SW intensities $S_{\xq \xq}^{(n)}(\delta )$
and $S_{\zq \zq }^{(n)}(\delta )$.  Hence, those modes excite spins within the $\xq \zq $ plane of the cycloid (neglecting its small tilt).  
On the other hand, $\Psi_n$ modes with $\langle \delta \vert M_{\xq } \vert 0 \rangle \ne 0$ or $\langle \delta \vert M_{\zq } \vert 0 \rangle \ne 0$
also have $S_{\yq \yq }^{(n)}(\delta ) > 0$.  Hence, those modes excite spins out of the $\xq \zq $ plane.  

Zone-center modes with nonzero MR matrix elements are indicated by the filled circles in Fig.3.  
In addition to having an enormous SW intensity, the ``zero"-frequency \cite{zero} $\Phi_0$ mode has a very large MR
matrix element (for $K=0.0035$ meV and $S_0=0.015$,  
$\vert \langle \delta \vert M_{\yq } \vert 0 \rangle \vert  \approx 8400 \mb $).  The $2\vQ $ potential splits
the degenerate $\Psi_{\pm 1}$ modes into $\Psi_1^{(1)} $ ($\langle \delta \vert M_{\xq } \vert 0 \rangle \ne 0$) and 
$\Psi_1^{(2)}$ ($\langle \delta \vert M_{\zq } \vert 0 \rangle \ne 0$).   The EM ($\langle \delta \vert P^{\rm ind}_{\yq }\vert 0\rangle \ne 0$) 
always coincides with $\Psi_1^{(1)}$.  Similarly, the smaller $4\vQ $ potential 
splits the $\Phi_{\pm 2}$ modes.  Due to its hybridization with $\Phi_0$, $\Phi_2^{(1)}$ becomes
spectroscopically active with $\langle \delta \vert M_{\yq } \vert 0 \rangle \ne 0$.

The predicted mode frequencies are plotted versus anisotropy for $S_0= 0.015$ in Fig.4.  
Both $\Psi_1^{(1,2)}$ and $\Phi_1^{(1,2)}$ cross near $K=0.001$ meV.  At $\eta = \delta $, $\Phi_1^{(2)}$ 
has no SW intensity and is not spectroscopically active.  But at $\eta = 0$, this mode is responsible for important 
features in the inelastic-scattering spectrum discussed in Section VI.  

For $K=0.0035$ meV, the mode frequencies are plotted versus $S_0$ in Fig.5(a), where $D$ and $D'$ are evaluated in terms of 
$S_0$ for fixed $\delta = 1/222$.  While the predicted spectroscopic mode frequencies decrease slightly with $S_0$, 
$\Phi_1^{(2)}$ slightly increases.

When $S_0=0$, the $\Psi_0$ and $\Phi_1^{(1)}$ modes at the zone center $\eta = \delta $
have no SW intensity and their MR matrix elements vanish.  But when $S_0 > 0$, the DM interaction ${\bf D'}$ 
with wavevector $(2\pi /a)[0.5,0.5,0.5]$ hybridizes $\Psi_0$ with $\Psi_1^{(1,2)}$ and $\Phi_1^{(1)}$ with $\Phi_0$.
Consequently, their MR matrix elements become significant.  

In Fig.5(b), the mode frequencies and MR matrix elements of $\Psi_0$ and $\Phi_1^{(1)}$
are plotted versus $S_0$ together with the very small SW intensities of those modes for $K=0.0035$ meV.  
As expected from perturbation theory, the matrix elements $\langle \delta \vert M_{\alpha } \vert 0 \rangle $ grow linearly
with $S_0 \sim \vert D'/J_1 \vert $.  Moreover, they scale like the square root of the SW intensities $S_{\alp \alp }(\delta )$.  
Therefore, these modes are both spectroscopically and dynamically activated by the tilt of the cycloid.
It is remarkable that the MR matrix elements of $\Psi_0$ and $\Phi_1^{(1)}$ become so large
while their SW intensities remain extremely weak. 

\begin{figure}
\includegraphics[width=8.5cm]{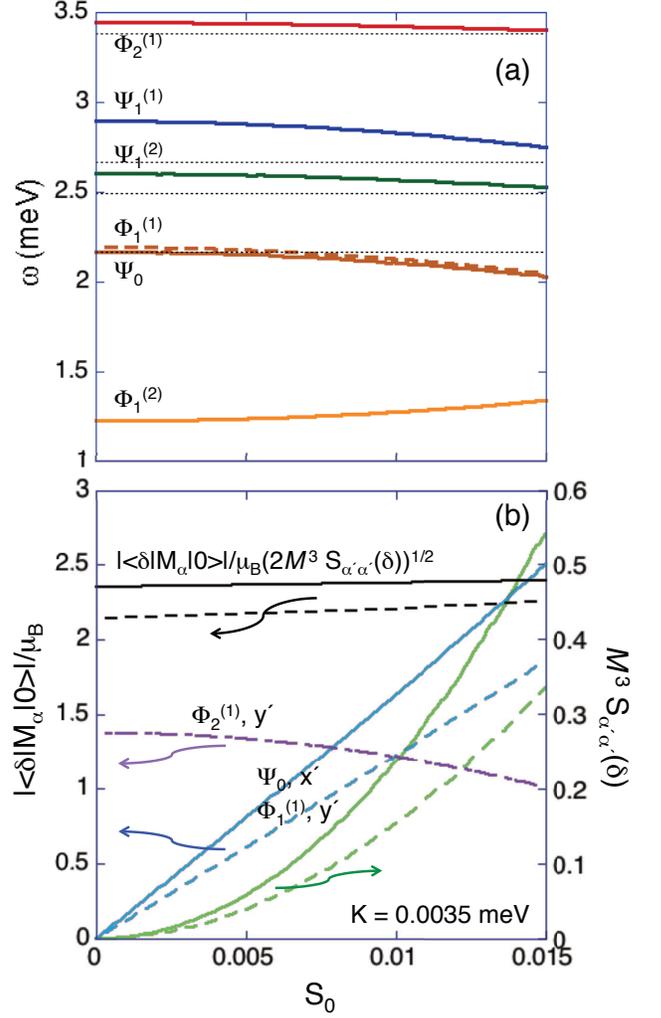}
\caption{(Color online) (a) The frequencies of the predicted modes versus $S_0$ for $K=0.0035$ meV.    
Horizontal dashed lines are the measured spectroscopic frequencies \cite{temp}.
(b) The MR matrix elements $|\langle  \delta | M_{\alpha } | 0\rangle |/\mu_B$ for 
$\Psi_0$ (solid) and $\Phi_1^{(1)}$ (dashed) versus $S_0$ for $K=0.0035$ meV.   
Also plotted are the intensities $M^3 S_{\alp \alp }(\delta )$ of those modes ($\alp = \yq$  for $\alpha = \xq$ and
$\alp = \xq$ or $\zq $ for $\alpha =\yq $) with $M= 222$.  The normalized matrix element 
$| \langle \delta | M_{\alpha } | 0\rangle |/\mu_B S_{\alp \alp }(\delta )^{1/2}$ is independent of $S_0$.  The dash-dot 
curve plots the MR matrix element for $\Phi_2^{(1)}$ with $\alpha = \yq $.}
\end{figure}

The dashed horizontal lines in Fig.4 correspond to the four measured spectroscopic frequencies of \BP.  
We believe that the nearly-degenerate $\Psi_0$ and $\Phi_1^{(1)}$ modes are responsible for the observed 
low-frequency peak at 2.17 meV.  Recall that those two modes only appear when the cycloid is tilted away from the 
$\xq \zq $ plane by the DM interaction ${\bf D'}$ along $\zp $.  The best overall fit to 
the observed mode spectrum is obtained with $K \approx 0.0035$ meV.  Measured \cite{temp} and 
predicted mode frequencies are summarized in Table I.

With $S_0 = 0.015$ and $K=0.0035$ meV, the harmonics of the cycloid have the ratio $C_3/C_1 = -0.050$ or $(C_1/C_3)^2 = 400$.  
Elastic neutron-scattering \cite{rama11a} and NMR measurements \cite{zalesskii02} indicate that $(C_1/C_3)^2 $ is 500 and 25, respectively.  However, 
the NMR measurement may overestimate the third harmonic due to the high $^{57}$Fe isotope content of the sample \cite{pokatilov10}.  
Our estimate for $(C_3/C_1)^2$ is in very good agreement with the elastic neutron-scattering result.

\vfill

\section{Selection Rules}

We now consider the selection rules for the THz modes \cite{talbayev11, nagelun}
for a sample with the single polarization domain $\vP =P\zp $, where $\zp =[1,1,1]$.  As mentioned in Section II, the three possible magnetic
domains have wavevectors $(2\pi /a)(0.5+\delta ,0.5, 0.5-\delta )$, $(2\pi /a) (0.5,0.5+\delta ,0.5-\delta )$,
and $(2\pi /a)(0.5+\delta ,0.5 -\delta ,0.5)$.  Since these domains have the same energy, we expect them to be equally populated.
The mode spectrum was measured for crossed fields $\vh_1 = [1,-1,0]$ and $\vh_2 = [1,1,0]$.   

To predict the selection rules for \BP , $\vh_1$ and $\vh_2$ are expressed in terms of the cycloidal unit vectors $\xp $, $\yp $, and $\zp $ as
\begin{eqnarray}
&&\vh_1=(\xp -\sqrt{3}\yp )/2,\nonumber \\
&&\vh_2= \xp /2 +\sqrt{3} \yp /6 +\sqrt{2/3}\zp ,
\end{eqnarray}
in domain 1 with $\xp = [1,0,-1]$ and $\yp = [-1,2,-1]$;
\begin{eqnarray}
&&\vh_1=-(\xp +\sqrt{3}\yp )/2, \nonumber \\
&&\vh_2= \xp /2 -\sqrt{3} \yp /6 +\sqrt{2/3}\zp ,
\end{eqnarray}
in domain 2 with $\xp = [0,1,-1]$ and $\yp = [-2,1,1]$; and
\begin{eqnarray}
&&\vh_1=\xp , \nonumber \\
&&\vh_2= (\yp +\sqrt{2}\zp )/\sqrt{3}, 
\end{eqnarray}
in domain 3 with $\xp = [1,-1,0]$ and $\yp = [1,1,-2]$.   
Although the following discussion assumes that all three domains are equally populated,
our qualitative conclusions remain unchanged even if one or two domain populations dominate the sample.

While $\Psi_1^{(1)}$ ($\langle \delta \vert M_{\xq } \vert 0 \rangle \ne 0$) and $\Phi_2^{(1)}$  
($\langle \delta \vert M_{\yq } \vert 0 \rangle \ne 0$) should appear in both fields $\vh_1$ and $\vh_2$, 
$\Psi_1^{(2)}$ ($\langle \delta \vert M_{\zq } \vert 0 \rangle \ne 0$) should only appear in field $\vh_2$,
which contains a $\zp $ component.
This agrees with the selection rule observed by Talbayev {\em et al.} [\onlinecite{talbayev11}].  But Nagel {\em at al.} [\onlinecite{nagelun}]
recently found that $\Psi_1^{(2)}$ survives in field $\vh_2$, although with drastically reduced intensity.  
Notice that the position of $\Psi_1^{(1)}$ above $\Psi_1^{(2)}$ requires that $K > 0.001$ meV.  Therefore, both nonzero 
$K$ and $S_0$ are required to explain the spectroscopic frequencies and selection rules.

Whereas Talbayev {\em et al.} \cite{talbayev11} found that the low-frequency mode appears only in field $\vh_1$, our
model indicates that the nearly-degenerate $\Psi_0 $ ($\langle \delta \vert M_{\xq } \vert 0 \rangle \ne 0$) and 
$\Phi_1^{(1)}$ ($\langle \delta \vert M_{\yq } \vert 0 \rangle \ne 0$) modes should appear in both fields $\vh_1$ and $\vh_2$.  
However, more precise THz measurements \cite{nagelun} have recently detected the low-frequency mode in both fields
$\vh_1$ and $\vh_2$.  At 4 K, Nagel {\em et al.} \cite{nagelun} even observed distinct low-frequency peaks
at 2.03 and 2.26 meV.  The observed three-fold splitting of the 2.03 meV peak in a magnetic field 
may help to distinguish $\Psi_0$ and $\Phi_1^{(1)}$.

To address the observability of the THz modes more carefully, we evaluate the spectroscopic intensities 
$I(\vh_1)$ and $I(\vh_2)$ for each mode.  The spectroscopic intensity for any mode is given by \cite{miyahara12}
\begin{equation}
\label{int}
I(\vh ) = \sum_{\alpha } h_{\alpha }^2 \, \vert \langle \delta \vert M_{\alpha } \vert 0\rangle \vert^2.
\end{equation}
Averaging over the three domains, we find 
\begin{eqnarray}
&&I(\vh_1) = \frac{1}{2} \Bigl\{ \vert \langle \delta \vert M_{\xq }\vert 0 \rangle \vert^2 + 
\vert \langle \delta \vert M_{\yq }\vert 0 \rangle \vert^2 \Bigr\} ,\\
&&I(\vh_2) = \frac{1}{6} \Bigl\{ \vert \langle \delta \vert M_{\xq }\vert 0 \rangle \vert^2 + 
\vert \langle \delta \vert M_{\yq }\vert 0 \rangle \vert^2 \Bigr\} \nonumber \\
&&+\frac{2}{3} \vert \langle \delta \vert M_{\zq }\vert 0 \rangle \vert^2 ,
\end{eqnarray}
For $\langle \delta \vert M_{\alpha }\vert 0 \rangle \ne 0$, $I(\vh_1)/I(\vh_2)=3$ for 
any mode (like $\Phi_2^{(1)}$, $\Psi_1^{(1)}$, $\Psi^{(0)}$, and $\Phi_1^{(1)}$)
with $\alpha  = \xq $ or $\yq $ while $I(\vh_1)/I(\vh_2)=0$ for any mode (like $\Psi_1^{(2)}$) with 
$\alpha = \zq $.  

\begin{table}
\caption{\textbf{Spectroscopic Frequencies, Matrix Elements, and Intensities}}
\begin{ruledtabular}
\begin{tabular}{lllllc}
   &  $\Psi_0$/$\Phi_1^{(1)}$& $\Psi_1^{(2)}$ & $\Psi_1^{(1)}$  & $\Phi_2^{(1)}$      \\
 \hline 
Measured $\omega \,$(meV) & 2.17 & 2.49 & 2.67 & 3.38  \\
Predicted $\omega \,$(meV) & 2.03/2.05 & 2.53 & 2.75 & 3.40 \\
MR index $\alpha $ & $\xq /\yq $ & $\zq $ & $\xq $ & $\yq $ \\ 
$\vert \langle \delta \vert M_{\alpha } \vert 0 \rangle \vert /\mb $ &2.50/1.86 & 3.96 & 4.59 &1.01  \\
$\vert \langle \delta \vert P_{\yq }\vert 0 \rangle \vert /\lambda $ & 0 & 0 & 12.2 & 0  \\
Intensity index $\alp $ & $\yq $/$\xq , \zq $ & $\yq $ & $\yq $& $\xq ,\zq $ \\
$S_{\alp \alp }(\delta )$& $4.94\times 10^{-8}$/  &  $19.7$ & $18.1$ & $5.43,$  \\
&$3.05\times 10^{-8}$ &&& 2.35  \\
$I(\vh_1)/\mb^2 $ & 4.75 & 0 & 10.54 & 0.51  \\
$I(\vh_2)/\mb^2 $ & 1.58 & 10.47 & 3.51 & 0.17  \\
\end{tabular}
\end{ruledtabular}
\end{table}

The spectroscopic intensities for $K=0.0035$ meV and $S_0=0.015$ are summarized in Table I.
These numerical results indicate that $\Psi_1^{(1)}$ and $\Psi_1^{(2)}$ should be the strongest
of the four modes, in agreement with the THz results \cite{talbayev11, nagelun}.  
Surprisingly, Table I indicates that the intensity $I(\vh_2)$ of $\Phi_2^{(1)}$ is roughly 20 times smaller than 
that of $\Psi_1^{(1)}$.  By contrast, recent THz measurements \cite{nagelun} indicate that $\Phi_2^{(1)}$ is only about 3 times
less intense than $\Psi_1^{(1)}$ in field $\vh_2$.  Those measurements do, however, agree with our prediction that $\Psi_1^{(2)}$ is 
several times more intense than $\Psi_1^{(1)}$ in $\vh_2$.

\section{Inelastic Neutron-Scattering Measurements}

In earlier work \cite{fishman12} with $D'=0$, we obtained conflicting estimates for the easy-axis anisotropy $K$ based on the spectroscopic and
neutron-scattering spectra.   Because the instrumental resolution is broader than $4\pi \delta /a$ [\onlinecite{matsuda12}], 
inelastic neutron-scattering measurements at the AF Bragg point $(2\pi /a)[0.5,0.5,0.5]$ average over a range of $\vq $ that includes 
both cycloidal satellites at $(2\pi /a)[0.5\pm \delta , 0.5, 0.5 \mp \delta ]$.  For $D'=0$, the spectroscopic mode frequencies indicated that 
$K \approx 0.002$ but the inelastic-scattering spectra indicated that $K \approx 0.004$.

\begin{figure}
\includegraphics[width=8.5cm]{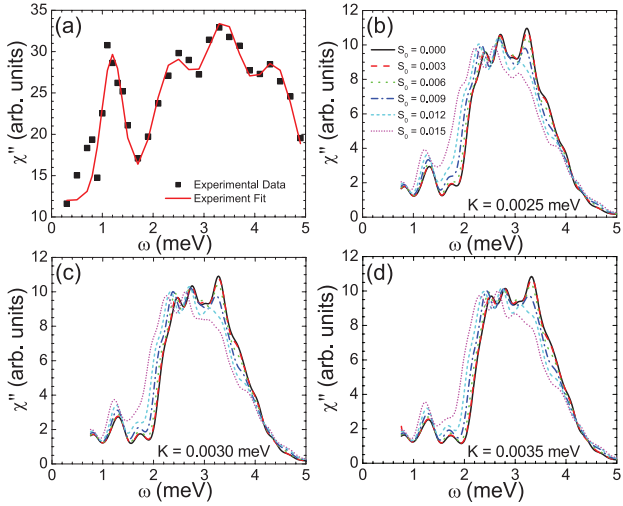}
\caption{(Color online) (a) The measured inelastic-scattering spectrum \cite{matsuda12, fishman12} around $\eta =0$ and
the predicted spectrum for (b) $K=0.0025$, (c) 0.003, and (d) 0.0035 meV with $S_0$ ranging from 0 to 0.015.
}
\end{figure}

We now re-examine the spectrum $\cdp (\omega )$ for $D' \ne 0$.  The upper left-hand corner of Fig.6 plots the 
measured spectrum \cite{matsuda12, fishman12}.  The resolution-averaged intensity spectrum is plotted versus $\omega $ in Figs.6(b-d) for 
three values of $K$ and six values of $S_0$ from 0 to 0.015.  The very low-frequency rise of $\cdp (\omega )$ due to 
$\Phi_0$ at $\eta = \delta $ has been removed from both the measured and predicted spectra.  

Below 5 meV, the measured $\cdp (\omega )$ contains four peaks at 1.2, 2.4, 3.4, and 4.4 meV.  
The peaks at 1.2 and 2.4 meV are primarily caused by $\Phi_1^{(1,2)}$ and $\Psi_0$.  
As shown in Fig.4 for $S_0=0.015$, the separation between $\Phi_1^{(2)}$ and $\Phi_1^{(1)}/\Psi_0$ increases as $K$ exceeds 0.001 meV.
Correspondingly, the gap in the predicted spectrum centered at 2 meV widens with increasing $K$ beyond 0.001 meV.

As shown in Fig.5(b), $\Phi_1^{(2)}$ is slightly enhanced by $S_0$.  But the resolution-averaged spectrum $\cdp (\omega )$ also involves nearby modes and 
shifts to lower frequencies with increasing $S_0$.  For $S_0=0.015$ and $K=0.0035$ meV, the low-frequency peak lies at 1.2 meV.  So 
based on this single peak, $K\approx 0.0035$ meV provides good agreement with both the spectroscopic and inelastic measurements.
Although its intensity increases with $S_0$ and it is more pronounced than in our previous work \cite{fishman12}, 
the predicted low-frequency peak at 1.2 meV is still considerably weaker than the measured peak.  

For $K=0.0035$ meV, the second peak lies at 2.5 meV when $S_0=0$ but shifts down to 2.3 meV when 
$S_0=0.015$.  More problematically, the predicted spectrum contains three peaks between 2 and 4 meV 
(although the third peak is suppressed with $S_0$) whereas the measured spectrum contains only two.  
For $K=0.0035$ meV and $S_0=0.015$, there are no predicted SW excitations between 4 and 5 meV at
$\eta =0 $ or $\delta $.  Consequently, the observed peak at 4.4 meV is missing from our spectrum, 
which falls off much more rapidly than the measured $\cdp (\omega )$ above 4 meV.  Keep in mind, however, 
that the predicted shape of $\cdp (\omega )$ sensitively depends on the resolution function used to perform the averaging.

\section{Conclusion}

A primary motivation of this work was to see how well a microscopic model can describe the properties of one of the simplest and
most technologically important multiferroic materials.  We have demonstrated that all four modes observed by THz and Raman spectroscopies in \BF
are predicted by a model that includes two DM interactions, one along $\yp $ responsible for the cycloid periodicity and the 
other along $\zp $ responsible for its tilt of the cycloid out of the
$\xq \zq $ plane.  Using reasonable values for the easy-axis anisotropy and the DM interactions, we obtain excellent agreement with the measured
mode frequencies.  The parameters $D= 0.11$ meV, $D'= 0.054$ meV, and $K=0.0035$ meV provide very good
descriptions of both the spectroscopic and inelastic neutron-scattering measurements, thereby resolving an earlier 
disagreement \cite{fishman12}.

The spectroscopic modes evolve with the complexity of the cycloid.  With a single DM interaction ${\bf D}=D\yp $, 
the cycloid is coplanar and purely harmonic.  For nonzero frequencies, the only spectroscopically-active mode is 
$\Psi_1$ ($\langle \delta \vert M_{\xq } \vert 0 \rangle \ne 0$,
$\langle \delta \vert M_{\zq} \vert 0 \rangle \ne 0$), which coincides with the EM ($\langle \delta \vert P^{{\rm ind}}_{\yq } \vert 0 \rangle \ne 0$).
Easy-axis anisotropy $K$ along $\zp $ distorts the coplanar cycloid and introduces higher even harmonics in the Hamiltonian $H$.  
The $2\vQ$ potential splits $\Psi_{\pm 1}$ into $\Psi_1^{(1)}$ ($\langle \delta \vert M_{\xq } \vert 0 \rangle \ne 0$, $\langle \delta \vert P^{{\rm ind}}_{\yq } \vert 0 \rangle \ne 0$)
and $\Psi_1^{(2)}$ ($\langle \delta \vert M_{\zq } \vert 0 \rangle \ne 0$);  the $4\vQ $ potential splits $\Phi_{\pm 2}$ into $\Phi_2^{(1)}$ and
$\Phi_2^{(2)}$.  Hybridized with $\Phi_0$ by the $2\vQ$ potential, $\Phi_2^{(1)}$ ($\langle \delta \vert M_{\yq } \vert 0 \rangle \ne 0$) becomes
spectroscopically active.  Finally, the DM interaction ${\bf D'}=D' \zp $ tilts the non-coplanar cycloid out of the $\xq \zq $ plane.  Then, 
$\Psi_0$ ($\langle \delta \vert M_{\xq } \vert 0 \rangle \ne 0$) and $\Phi_1^{(1)}$ ($\langle \delta \vert M_{\yq } \vert 0 \rangle \ne 0$) 
are dynamically and spectroscopically activated by their hybridization with $\Psi_1^{(1,2)}$ and $\Phi_0$, respectively.  
Thus, additional interactions modify the mode spectrum as more modes hybridize with $\Phi_0$ and $\Psi_1^{(1,2)}$.

Several experiments indicate that the low-temperature, low-field cycloid of \BF undergoes a transition
at about 140 K or 10 T.  In THz measurements \cite{talbayev11}, the low-frequency $\Psi_0$/$\Phi_1^{(1)}$ mode disappears above 120 K
and the high-frequency $\Phi_2^{(1)}$ mode disappears above 150 K.  
Nevertheless, the selection rules governing the $\Psi_1^{(1,2)}$ modes do not change \cite{talbayev11}.  
In Raman measurements, all modes persist for all temperatures but their frequencies \cite{cazayous08} and intensities \cite{singh08} display kinks 
at about 140 K.  Optical \cite{xu09} and electron-spin resonance \cite{ruette04} measurements show anomalies at about 10 T with indications
that the cycloidal phase above 10 T is the same as the one above 140 K.  
Recently, Nagel {\em et al.} [\onlinecite{nagelun}] found that the THz modes exhibit kinks at about 5.5 T.
But the nature of these transitions and the difference between the two cycloidal phases remain unknown.

With magnetic field along $\zp $, the Hamiltonian of Eq.(\ref{Ham}) does not produce a transition between 
different cycloidal phases \cite{unpub}.  Therefore, the proposed model may be incomplete.  
Since $D'$ is responsible for the low-frequency $\Psi_0$/$\Phi_1^{(1)}$ mode, a sudden change in 
$D'$ at 140 K or 10 T would produce anomalies in its spectroscopic features.  A jump in $D'$ at 140 K would also produce
a jump in the weak ferromagnetic moment $M_0(T)$.  We hope that future experimental and theoretical work will 
resolve this and other mysteries surrounding \BP .

We gratefully acknowledge conversations with Masaaki Matsuda, Jan Musfeldt, Satoshi Okamoto,
and Toomas R\~o\~on.
Research sponsored by the U.S. Department of Energy, Office of Basic Energy Sciences, 
Materials Sciences and Engineering Division (RF), by the Center for Integrated Nanotechnologies, a 
U.S. Department of Energy, Office of Basic Energy Sciences user facility at Los Alamos National Laboratory, operated by 
Los Alamos National Security, LLC for the National Nuclear Security Administration of the U.S. Department of Energy (JH),
by Grants-in-Aid for Scientific Research from the Ministry of Education, Culture, and Technology, Japan (MEXT) (NF),
and by the Max Planck Society (MPG), the Korea Ministry of Education, Science and Technology (MEST),
Gyeongsangbuk-Do and Pohang City (SM).

\appendix

\section{SW intensities}

This section describes how to evaluate the SW intensities and eigenvectors $\uX $, which are required in the 
next section to evaluate the spectroscopic matrix elements.

The local reference frame for each spin $\vS_i$ on site $i$ is defined in terms of the unitary matrix $\underline{U}^i$
by ${\bf {\bar S}}_i = \underline{U}^i \, \vS_i$.  For spin 
\begin{equation}
\vS  = S\bigl(\sin \theta \cos \phi , \sin \theta \sin \phi , \cos \theta \bigr),
\end{equation}
the matrices $\underline{U}$ and $\underline{U}^{-1}$ are given by
\begin{equation}
\underline{U}=\left( \begin{array}{ccc}\cos \theta \cos \phi & \cos \theta \sin \phi & -\sin \theta \\
-\sin \phi & \cos \phi & 0 \\
\sin \theta \cos \phi & \sin \theta \sin \phi & \cos \theta 
\end{array} \right), 
\end{equation}
\begin{equation}
\underline{U}^{-1}=\left( \begin{array}{ccc}\cos \theta \cos \phi & -\sin \phi & \sin \theta \cos \phi \\
\cos \theta \sin \phi & \cos \phi & \sin \theta \sin \phi \\
-\sin \theta & 0 & \cos \theta 
\end{array} \right),
\end{equation}
so that $S\underline{U}^{-1}\cdot {\bf z} = \vS $.  

A Holstein-Primakoff transformation is used to express the local spin operators ${\bf {\bar S}}_i$ in terms of the 
bosons $a_i$ and $a_i^{\dagger }$ with $\bar{S}_{iz} = S -a_i^{\dagger }a_i$, 
$\bar{S}_{i+} = \sqrt{2S} a_i$, and $\bar{S}_{i-}=\sqrt{2S} a_i^{\dagger }$.
The Hamiltonian is then expanded in powers of $1/\sqrt{S}$ as $H=E_0 + H_1 +H_2+\ldots $.  While
$E_0$ is the classical energy and $H_1$ must vanish, 
\begin{equation}
H_2 =\sum_{\vq } \vv^{\dagger}_{\vq } \cdot \underline{L}(\vq )\cdot \vv_{\vq },
\end{equation}
where $\vv_{\vq }=(a_{\vq }^{(1)},\ldots ,a_{\vq }^{(2M)},a_{-\vq }^{(1)\dagger },\ldots ,a_{-\vq }^{(2M)\dagger })$ is a $4M$-dimensional
vector and $\underline{L}(\vq )$ is a $4M$-dimensional matrix.   Boson operators $a_{\vq }^{(r)}$ with $1 \le r \le M=222$ reside on layer
1 of the unit cell while those with $M+1 \le r \le 2M $ reside on layer 2.   The sublattice index $r$ refers to sites on either layer
with $\vR \cdot \xp =[r]a/\sqrt{2} $ where $[r]\equiv {\rm mod}(r,M)$.

Since $a_{\vq }^{(r)}$ and $a_{\vq }^{(r) \dagger }$
obey the commutation relations $[a_{\vq }^{(r)}, a_{\vq'}^{(s)\dagger }]=\delta_{r,s}\delta_{\vq ,\vq'}$ and
$[a_{\vq }^{(r)}, a_{\vq'}^{(s)}]=0$, $\vv_{\vq }$ and $\vv_{\vq }^{\dagger }$
satisfy the commutation relation $[\vv_{\vq },\vv^{\dagger }_{\vq'}]=\underline{N}\delta_{\vq ,\vq'}$ where
\begin{equation}
\underline{N}=\left( \begin{array}{cc}\underline{I} & 0 \\
0 & -\underline{I} 
\end{array} \right) 
\end{equation}
and $\underline{I}$ is the $2M$-dimensional unit matrix.

A diagonal form for $H_2$ is given by
\begin{equation}
H_2 =\sum_{\vq } \ww^{\dagger}_{\vq } \cdot \underline{L'}(\vq )\cdot \ww_{\vq },
\end{equation}
where $\ww_{\vq }=(\alpha_{\vq }^{(1)},\ldots ,\alpha_{\vq }^{(2M)},\alpha_{-\vq }^{(1)\dagger },\ldots ,\alpha_{-\vq }^{(2M)\dagger })$
and the boson operators $\alpha_{\vq }^{(n)}$ and $\alpha_{\vq }^{(n)\dagger }$ also obey canonical commutation relations.
The $4M$-dimensional matrix $\underline{L'}(\vq )$ is diagonal with real eigenvalues 
$\epsilon_n (\vq ) = \omega_n ( \vq )/2 > 0$ ($n=1,\ldots ,2M$)
and $\epsilon_n (\vq ) = -\omega_n (\vq )/2 < 0$ ($n=2M+1,\ldots ,4M$).  So for each $\vq $,
there are $2M$ positive and $2M$ negative eigenvalues.  The commutation relations yield
\begin{equation}
H_2 =\sum_{n,\, \vk } \omega_n (\vq ) \biggl\{ \alpha_{\vq }^{(n)\dagger }\alpha_{\vq }^{(n)} + \frac{1}{2} \biggr\},
\end{equation}
which identifies $\omega_n (\vq )$ as the SW frequency for mode $n$ with wavevector $\vq $.

Vectors $\ww_{\vq }$ and $\vv_{\vq }$ are related by
$\ww_{\vq }=\underline{X}(\vq )\cdot \vv_{\vq }$ or $\vv_{\vq }=\underline{X}^{-1}(\vq )\cdot \ww_{\vq }$, 
where the $4M$-dimensional matrix $\underline{X}$ is normalized by 
$\underline{X}\cdot \underline{N}\cdot \underline{X}^{\dagger } = \underline{N}$.  For fixed $\vq $, 
\begin{equation}
\sum_j \Bigl( {\cal{L}}_{ij}(\vq )- \delta_{ij}\epsilon_n(\vq ) \Bigr) X^*_{nj}(\vq ) = 0,
\end{equation}
where $\underline{\cal{L}}(\vq )=\underline{L}(\vq )\cdot {\underline N}$.
The inverse $\underline{X}^{-1}=\underline{N} \cdot \underline{X}^{\dagger }\cdot \underline{N}$ 
is required to evaluate $\langle \delta \vert \pin \vert 0\rangle $ and $\langle \delta \vert {\bf M}\vert 0\rangle $.

The wavevector $\vQ $ and harmonic coefficients of the cycloid are obtained by minimizing $E_0$ using
the ``trial" spin state provided by Eqs.(\ref{dcx}-\ref{dcz}).  If the spin angles on site $r$ of layer 1 are $\theta_r$
and $\phi_r$, then the angles on layers 1 and 2 are related by 
$\theta_{r+M} = \theta_r + \pi $ and $\phi_{r+M} = -\phi_r$.  
We assume that $\phi_r = \tau $ and $\phi_{r+M}=-\tau $ are independent of site position $r$ on layers 1 and 2.

The spin-spin correlation function is defined by
\begin{eqnarray}
\label{ssc}
S_{\alpha \beta }(\vq ,\omega ) &=& \frac{1}{2\pi N} \int dt \, e^{-i\omega t} \sum_{i,j} e^{-i\vq \cdot (\vR_i -\vR_j )}
\nonumber \\ &&\langle S_{i\alpha }(0)S_{j\beta }(t)\rangle \nonumber \\
&=&\sum_n \delta \bigl( \omega - \omega_n (\vq ) \bigr) S_{\alpha \beta }^{(n)}(\vq ),
\end{eqnarray}
where the final expression assumes that the SWs are undamped.  
The inelastic neutron-scattering cross section is \cite{shirane04}
\begin{eqnarray}
\label{sqw}
&&S(\vq ,\omega ) =\sum_{\alpha , \beta }\Bigl( \delta_{\alpha \beta } - q_{\alpha }q_{\beta }/q^2 \Bigr) S_{\alpha \beta }(\vq ,\omega ) \nonumber \\
&& = \sum_{n, \alpha } \Bigl(1- (q_{\alpha }/q)^2 \Bigr) \delta \bigl( \omega - \omega_n(\vq ) \bigr) S^{(n)}_{\alpha \alpha }(\vq ), \,\,\,\,\,\,\,\,\,
\end{eqnarray}
which only involves the diagonal matrix elements of $S_{\alpha \beta }(\vq ,\omega )$ (if there is a net moment, some 
off-diagonal matrix elements $\alpha \ne \beta $ are nonzero and antisymmetric). 
The diagonal SW intensities $S_{\alpha \alpha }^{(n)}(\vq )$ are given by
\begin{equation}
\label{sec}
S_{\alpha \alpha }^{(n)}(\vq ) = \frac{S}{8M} \sum_{r=1}^{2M} \Bigl\vert W_{r ,\alpha }^{(n)}(\vq ) \Bigr\vert^2 ,
\end{equation}
where
\begin{eqnarray}
&&W_{r ,\alpha }^{(n)}(\vq ) = 
\bigl(\Ui_{\alpha x} -i\Ui_{\alpha y} \bigr) X^{-1}_{r,n+2M}(\vq ) \nonumber \\
&&+\bigl( \Ui_{\alpha x}+i\Ui_{\alpha y} \bigr) X^{-1}_{r+2M,n+2M}(\vq ).
\end{eqnarray}
Even in the absence of damping, the instrumental resolution will broaden the delta functions in $S(\vq ,\omega )$ in Eq.(\ref{sqw}).
The magnetic form factor for Fe$^{3+}$ should also be included in $S(\vq , \omega )$.

\section{Spectroscopic matrix elements}

This section evaluates the matrix elements for the induced electric polarization $\pin $ and the
magnetic moment ${\bf M}$ between the ground state $\vert 0 \rangle $ and an excited state $\vert \delta \rangle $
with a single magnon at the cycloidal wavevector $\vQ $.

Since $P^{\rm ind}_{\xq }=0$, only the $\yq $ and $\zq $ 
components are considered.  Expanded about equilibrium, $P^{\rm ind}_{\yq }$ becomes
\begin{eqnarray}
&&P^{\rm ind}_{\yq }=\lambda S\Biggl\{ \sum_{r=1}^M \sin \theta_r \cos \phi_r \,
\Bigl[ -S_{[r+2], \yq }+S_{[r-2], \yq } \nonumber \\
&&+S_{[r+2]+M, \yq }-S_{[r-2]+M ,\yq }\Bigr] \nonumber \\
&&+ \ds\sum_{r=1}^M \sin \theta_r \sin \phi_r\, \Bigl[ S_{[r+2], \xq }- S_{[r-2], \xq }
\nonumber \\
&&+S_{[r+2]+M, \xq }-S_{[r-2]+M ,\xq }\Bigr] \Biggr\}.
\end{eqnarray}
After some work, we obtain the EM matrix element $\yq $ for SW mode $n$: 
\begin{eqnarray}
\label{emme}
&&\langle \delta \vert P^{\rm ind}_{\yq } \vert 0 \rangle  =\lambda S \sqrt{\frac{ S}{2}}\,
\sum_{r=1}^M \sin \theta_r \, e^{i q_0 a r  }
\nonumber \\ 
&&\biggl\{ \Bigl[ \cos \theta_{[r+2]} \sin (\phi_r -\phi_{[r+2] })+ i \cos (\phi_r -\phi_{[r+2]} ) \Bigl] \nonumber \\
&&\Bigl( X^{-1}_{[r+2],n+2M} - X^{-1}_{[r+2]+M,n+2M} 
\Bigr) e^{2i q_0 a  }\nonumber \\
&&+\Bigl[ \cos \theta_{[r+2]} \sin (\phi_r -\phi_{[r+2] }) - i \cos (\phi_r -\phi_{[r+2]}) \Bigr] \nonumber \\
&&\Bigl( X^{-1}_{[r+2]+2M,n+2M} - X^{-1}_{[r+2]+3M,n+2M} 
\Bigr) e^{2iq_0 a  } \nonumber \\
&&-\Bigl[ \cos \theta_{[r-2]} \sin (\phi_r -\phi_{[r-2] }) + i \cos (\phi_r -\phi_{[r-2]}) \Bigr] \nonumber \\
&&\Bigl( X^{-1}_{[r-2],n+2M} - X^{-1}_{[r-2]+M,n+2M} 
\Bigr) e^{-2iq_0 a } \nonumber \\
&&-\Bigl[ \cos \theta_{[r-2]} \sin (\phi_r -\phi_{[r-2] }) - i \cos (\phi_r -\phi_{[r-2]}) \Bigr] \nonumber \\
&&\Bigl( X^{-1}_{[r-2]+2M,n+2M} - X^{-1}_{[r-2]+3M,n+2M} 
\Bigr) \nonumber \\
&& e^{-2iq_0 a  }
\biggr\} , 
\end{eqnarray}
where $q_0 = 2\pi \delta /a$.  

Similarly, $P^{\rm ind}_{\zq }$ can be expanded as
\begin{eqnarray}
&&P^{\rm ind}_{\zq }=\lambda S\Biggl\{  \sum_{r=1}^M \cos \theta_r \,
\Bigl[ S_{[r+2], \xq }-S_{[r-2], \xq } \nonumber \\
&& -S_{[r+2]+M, \xq }+S_{[r-2]+M, \xq } \Bigr] \nonumber \\
&&-\ds\sum_{r=1}^M \sin \theta_r \cos \phi_r\, \Bigl[ 
S_{[r+2], \zq }- S_{[r-2], \zq } \nonumber \\
&&-S_{[r+2]+M, \zq }+ S_{[r-2]+M, \zq } \Bigr] \Biggr\}.
\end{eqnarray}
The EM matrix element $\zq $ for SW mode $n$ is 
\begin{eqnarray}
&&\langle \delta \vert P^{\rm ind}_{\zq } \vert 0 \rangle  =\lambda S \sqrt{\frac{ S}{2}}\, 
\sum_{r=1}^M e^{i q_0 ar} 
\nonumber \\ 
&&\biggl\{ 
\Bigl[ g_{r,[r+2]} + i \cos \theta_r \sin \phi_{[r+2]} \Bigl]  \Bigl( X^{-1}_{[r+2],n+2M} 
\nonumber \\
&&-X^{-1}_{[r+2]+M,n+2M} \Bigr) e^{2iq_0 a}\nonumber \\
&&+\Bigl[ g_{r,[r+2]} - i \cos \theta_r \sin \phi_{[r+2]} \Bigr]  
\Bigl( X^{-1}_{[r+2]+2M,n+2M} \nonumber \\
&&- X^{-1}_{[r+2]+3M,n+2M} \Bigr) e^{2iq_0 a}\nonumber \\
&& -\Bigl[ g_{r,[r-2]} + i \cos \theta_r \sin \phi_{[r+2]} \Bigl]  \Bigl( X^{-1}_{[r-2],n+2M} 
\nonumber \\
&&-X^{-1}_{[r-2]+M,n+2M} \Bigr) e^{-2iq_0 a }\nonumber \\
&&-\Bigl[ g_{r,[r+2]} - i \cos \theta_r \sin \phi_{[r-2]} \Bigr]  \Bigl( X^{-1}_{[r-2]+2M,n+2M} 
\nonumber \\
&&- X^{-1}_{[r-2]+3M,n+2M} \Bigr) e^{-2iq_0 a } \biggr\} , 
\end{eqnarray}
where
\begin{equation}
g_{r,s} = \cos \theta_r \cos \theta_s \cos \phi_s + \sin \theta_r \sin \theta_s \cos \phi_r .
\end{equation}
For $K=0.0035$ meV and $S_0=0.015$,
$\Phi_0$ has the small matrix element $\langle \delta \vert P^{\rm ind}_{\zq }\vert 0\rangle \approx 0.19$,
about 60 times smaller than $\langle \delta \vert P^{\rm ind}_{\yq }\vert 0\rangle \approx 12.2$ for $\Psi_1^{(1)}$.

The MR matrix element for SW mode $n$ is much more simply given by
\begin{eqnarray}
\label{mrme}
&&\langle \delta \vert M_{\alpha  } \vert 0 \rangle =  \sqrt{2S}\mb \,
\sum_{r=1}^{2M} e^{i q_0 a [r] } \nonumber \\ 
&&\sgn (M-r+1/2)\, W_{r ,\alpha }^{(n)}(\vQ ) ,
\end{eqnarray}
which uses 
\begin{equation}
e^{i \vQ \cdot \vR } =e^{i q_0 a [r]} \, \sgn (M-r+1/2).
\end{equation}
Notice that $W_{r ,\alpha }^{(n)}(\vq )$ also enters the SW intensity $S_{\alpha \alpha }^{(n)}(\vq )$ of Eq.(\ref{sec}).  While the SW intensity 
$S_{\alpha \alpha }^{(n)}(\vQ )$ is proportional to the sum of $\vert W_{r ,\alpha }^{(n)}(\vQ)\vert^2$ over $r$, 
the matrix element $\langle \delta \vert M_{\alpha  } \vert 0 \rangle $ is proportional to the Fourier transform 
of $W_{r ,\alpha }^{(n)}(\vQ) $ over $r$.


\begin{thebibliography}{}
\bibitem{temp} To account for the suppression of the observed moment at 200 K,
Ref.[\onlinecite{matsuda12}] took $2\mu_B\sqrt{S(S+1)}= 4.11\, \mu_B$ to be about 30\% lower than its value with $S=5/2$.
Here we take $2\mu_B\sqrt{S(S+1)}=5.92\, \mu_B$
and reduce the exchange interactions accordingly.  So the estimates $J_1=6.48$ meV, $J_2=0.29$ meV, and $D=0.162$ meV from
Ref.[\onlinecite{matsuda12}] are equivalent to $J_1=-4.5$ meV, $J_2=-0.2$ meV, and $D=0.119$ meV in this paper.
All spectroscopic frequencies are measured at or extrapolated to 200 K.
\bibitem{zero} $\Phi_0$ is not stricty a Goldstone mode because rotational invariance about $\zp $ is broken by 
${\bf D}$.  However, the energy gap is so small that we are unable to reliably estimate its size.
\end{thebibliography}
\end{document}